\documentclass[a4paper,11pt]{article}
\usepackage{pos}
\usepackage{diagbox}
\usepackage{bbm}
\usepackage{amsmath}
\usepackage{graphicx}
\usepackage{subcaption}
\usepackage{fdsymbol}

\PassOptionsToPackage{hyphens}{url}
\usepackage{hyperref}
\usepackage{color}

\title{
{\normalsize {\rm \hfill{HU-EP-22/42-RTG} }}\\
\bigskip
\bigskip
\bigskip
Supersphere non-linear sigma model on the lattice 
}

\author*{Ilaria Costa$^{\vardiamondsuit}$}
\author{Valentina Forini$^{\spadesuit, \vardiamondsuit}$, Ben Hoare$^{\clubsuit}$, Tim Meier$^{\vardiamondsuit}$, Agostino Patella$^{\vardiamondsuit}$, \\Johannes H. Weber$^{\vardiamondsuit}$}

\vspace{0.3cm}

\affiliation{$^\vardiamondsuit$Institut f\"ur Physik, Humboldt-Universit\"at zu Berlin, IRIS Adlershof, Berlin
\\$^\spadesuit$Department of Mathematics, City University London, Northampton Square, London ECIV 0HB, UK
\\$^\clubsuit$Department of Mathematical Sciences, Durham University, Durham DH1 3LE, UK}

\vspace{0.3cm}

\emailAdd{ilaria.costa@physik.hu-berlin.de}
\emailAdd{valentina.forini@city.ac.uk}
\emailAdd{agostino.patella@physik.hu-berlin.de}
\emailAdd{ben.hoare@durham.ac.uk}
\emailAdd{tmeier@physik.hu-berlin.de}
\emailAdd{johannes.weber@physik.hu-berlin.de}
\FullConference{39th International Symposium on Lattice Field Theory \\ 08--13 August 2022 \\ University of Bonn}

\abstract{
Two-dimensional $O(N)$ non-linear sigma models are exactly solvable theories and have many applications, from statistical mechanics to their use as QCD toy models. We consider a supersymmetric extension,  the non-linear sigma model on the supersphere~$S^{N+2m-1|2m}\equiv \frac{OSP(N+2m|2m)}{OSP(N+2m-1|2m)}$. We briefly describe its renormalization properties and lattice discretization, and present a strategy for numerical simulations together with some preliminary numerical results.
}
\begin{document}
\maketitle

\section{Introduction}



Under certain conditions,  nonlinear sigma models (NLSM) are renormalizable and even completely solvable. An example is the $O(N)$ NLSM: in two dimensions the model is renormalizable~\cite{Brezin1976} and appears in a variety of contexts in statistical mechanics as well as a QCD toy model~\cite{Novikov:1984ac,DAdda:1978dle,Polyakov:1975rr,Pelissetto:2000ek} and consequently has been an object of thorough study via lattice QFT methods. Its quantum integrability has been demonstrated in~\cite{Zamolodchikov1979} by showing the factorization of the S-matrix.  A simple supersymmetric extension of the $O(N)$ NLSM - with target space supersymmetry~\footnote{A well known different type of supersymmetric extension is the one which considers a supersymmetric worldsheet~\cite{Witten:1977xn, DiVecchia:1977nxl}. For its lattice QFT analysis see for example~\cite{Flore:2012xj}.} --
is the sigma model with target space $OSP(N+2m|2m)/OSP(N+2m-1|2m)\equiv S^{N+2m-1|2m}$,  a supersphere.  
Some analytic properties of this model such as the spectrum of local operators at the renormalization group fixed-points, their integrability properties and their integrable deformations have been studied in~\cite{Read:2001pz, Saleur:2001cw, Saleur:2003zm, Babichenko:2006uc,Mitev:2008yt, Cagnazzo:2014yha, Alfimov:2020jpy}. 
%
Below we sketch the renormalization properties and  the lattice discretization of this model,  presenting some preliminary, standard Hybrid Monte Carlo numerics for fermionic two-point functions and effective masses. %
This study, whose details will be given in a separate publication, provides the simplest ground where to gain experience on the lattice QFT analysis of two-dimensional sigma models on supersymmetric target spaces.  The latter play a role in a variety of models in statistical mechanics~\cite{Parisi:1980in, Gruzberg:1999dk, Quella:2013oda, Zirnbauer:2018ooz} and, notably, in string theory and the AdS/CFT correspondence~\cite{Maldacena:1997re,Metsaev:1998it} (see~\cite{Forini:2016sot,Forini:2016gie,Bianchi:2016cyv, Bianchi:2019ygz, Bliard:2022kne,Bliard:2022oof} for an account of the challenges underlying the discretization of gauge-fixed worldsheet models).


\section{The Model}

We consider a 2-dimensional NLSM whose target space is the supersphere $S^{N+2m-1|2m}\equiv OSP(N+2m|2m)/OSP(N+2m-1|2m)$. Consider on $\mathbb{R}^{N+2m|2m}$ a multiplet of supercoordinates~
$\Phi\equiv(\xi^1,\ldots,\xi^{N+2m},\psi^1,\ldots,\psi^{2m})$, where $\xi^a$ and $\psi^\alpha$ represent commuting (bosonic) and anticommuting (fermionic) degrees of freedom respectively. For two such multiplets one can define an inner product 
\begin{equation}
\Phi\cdot\tilde\Phi=\xi^a\tilde\xi^a+\mathcal{J}^{\alpha\beta}\psi_\alpha\tilde\psi_\beta\,,
\end{equation}
where repeated indices are summed and $\mathcal{J}_{\alpha\beta}$ is the $2m\times2m$-dimensional canonical symplectic matrix
\begin{equation}
	\mathcal{J}_{\alpha\beta}=\left(\begin{array}{cc}
	0& \mathbbm{1}\\-\mathbbm{1}&0
	\end{array}\right).
\end{equation}
The unit supersphere constraint is defined by 
\begin{equation} 
\Phi\cdot\Phi=\xi^a\xi^a+\mathcal{J}^{\alpha\beta}\psi_\alpha\psi_\beta=1\,.
\end{equation}
In the lattice NLSM,  coordinates on the supersphere are promoted to lattice-discretized fields $\Phi_x$ (with mixed bosonic and fermionic coordinates) and the lattice-discretized path integral is defined as
\begin{equation}
\mathcal{Z}=\int\,\mathcal{D}\Phi\, e^{-\mathcal{S}_0},
\label{path_int}
\end{equation}
where the action and measure are
\begin{equation}
\begin{split}
\mathcal{S}_0&=\frac{a^2}{g}\sum_{x}\partial^f_{\mu} \Phi_x \cdot \partial^f_{\mu} \Phi_x
	=\frac{a^2}{g}\sum_{x}\left[2-\mathcal{J}^{\alpha\beta}\psi^{\alpha}_{x+\mu}\psi^{\beta}_x-2\xi^{a}_{x+\mu}\xi^{a}_x\right],\\
\mathcal{D}\Phi &=\prod_{x}\delta\left(1-\xi_x^a\xi_x^a-\mathcal J^{\alpha\beta}\psi^\alpha_x\psi^\beta_x\right) d\xi_x d\psi_x.
\end{split}
\label{inv_meas}
\end{equation}
Both action and  measure are invariant under the supergroup $OSP(N+2m|2m)$, whose algebra can be represented by the super-matrix
\begin{equation}
S=	\left(\begin{array}{cc}
	S_{\xi \xi} & S_{\xi \psi} \\
	S_{\psi \xi} & S_{\psi \psi}
	\end{array}\right), 
\end{equation}
where $S_{\xi\xi}$ is an element of the $\mathfrak{so}(N+2m)$ algebra, $S_{\psi\psi}\in \mathfrak{sp}(2m,\mathbb{R})$, while $S_{\xi\psi}$ and $S_{\psi\xi}$ are anticommuting $\left(N+2m\right)\times2m$ and $2m\times \left(N+2m\right)$-dimensional matrices respectively, satisfying the condition $S_{\xi\psi}=-S_{\psi\xi}^T\,\mathcal{J}$. The field coordinates transform as
$\delta\Phi=S\,\Phi$, or explicitly $\delta \xi=S_{\xi\xi}\xi+S_{\xi\psi}\psi$, $\delta \psi=S_{\psi\xi}\xi+S_{\psi\psi}\psi$. 

For $m=0$ and $N=1$ the  supersphere NLSM reduces to the Ising model.  For all other cases, we verified that the models are renormalizable at all orders in perturbation theory, both in dimensional regularization and on the lattice. 
To show this, it is sufficient to generalize the steps of the purely bosonic case~\cite{Brezin1976}. The non-linear realization of the $OSP(N+2m|2m)$ symmetry has strong implications on the form of divergences in perturbation theory.  
The Ward-Takahashi identities constrain the form of possible counterterms, whose coefficients can be calculated as a function of only two renormalization constants - the coupling constant $Z_g$ and a unique field renormalization $Z_\Phi$.   
The full account of this procedure will be given in a separate publication. 

\section{Formulation with auxiliary fields}
From now on we restrict to the case $m=1$, i.e. with only two fermionic degrees of freedom. 
In order to find a form of the action amenable to numerical simulations for this theory, we need to integrate out the two fermionic fields and in order to do so, we first need to get rid of them in the constraint in \eqref{inv_meas}. This can be done introducing a coordinate change for the bosonic fields
\begin{equation}
\xi^{a}_x=\rho_x\varphi^a_x \qquad \text{with}~~~~~~~ \varphi^2_x=1\qquad\text{and}\qquad d\xi_x=\rho^{N+1}_xd\rho_x d\varphi_x {\,\,\delta(\varphi^2_x-1)}.
\end{equation}
The path integral assumes the form
\begin{equation}
\mathcal{Z}=\int\,\prod_{x}\,d\rho_x d\varphi_x d\psi_x\,\delta(\rho^2+2\psi_1\psi_2\,\mathcal J_{12}-1){\delta(\varphi^2_x-1)}\rho^{N+1}_x\,e^{-\mathcal{S}_0}\,,
\end{equation}
where the integral over the $\rho$ field is limited to the interval $\left[0,\infty\right)$.
\noindent Integrating out the field $\rho$ one obtains
\begin{equation}
\begin{split}
\mathcal{Z}=\int \,\prod_{x}\,d\varphi_x d\psi_x\,\delta(\varphi^2_x-1)\left(1-N\psi^1_x\psi^2_x\,\mathcal J_{12}\right)\,e^{-\mathcal{S}_0}
=\int\,\prod_{x}\, d\varphi_x d\psi_x\,\delta(\varphi^2_x-1)e^{-\mathcal{S}_1},
\end{split}
\label{pth_int_2}
\end{equation}
where
\begin{equation}
\begin{split}
\mathcal{S}_1&=\sum_x\,N\,\psi^1_x\psi^2_x\,\mathcal J_{12}+\frac{2}{g}\sum_{x}\left[1-\varphi^a_{x+\mu}\varphi^a_{x}+\psi^1_{x}\psi^2_{x}\varphi^a_{x}\left(\varphi^a_{x+\mu}+\varphi^a_{x-\mu}\right)\mathcal J_{12}\right.\\&\left.-\psi^1_{x+\mu}\psi^2_{x+\mu}\psi^1_{x}\psi^2_{x}\varphi^a_{x+\mu}\varphi^a_x\mathcal{J}^2_{12}-\psi^1_{x}\left(\psi^2_{x+\mu}+\psi^2_{x-\mu}\right)\mathcal{J}_{12}\right].
\end{split}
\end{equation}
Notice the presence of a four-fermion, two-boson interaction term.

\noindent Completing the square $\psi^1_{x+\mu}\psi^2_{x+\mu}\psi^1_{x}\psi^2_{x}\varphi^a_{x+\mu}\varphi^a_x=\frac{1}{2}(\psi^1_{x+\mu}\psi^2_{x+\mu}\varphi^a_{x+\mu}+\psi^1_{x}\psi^2_{x}\varphi^a_{x})^2$, we can further manipulate the path integral by applying the  Hubbard-Stratonovich transformation 
\begin{equation}
e^{\frac{\zeta^2}{g}}=\sqrt{g\pi}\int\,dA\,e^{-\frac{1}{g}(A^2+\,2A\zeta)}\quad \text{with  }\zeta_{x,\mu,n}=\left(\varphi^a_{x+\mu}\psi^1_{x+\mu}\psi^2_{x+\mu}+\varphi^a_{x}\psi^1_{x}\psi^2_{x}\right)\mathcal{J}_{12}.
\end{equation}
for every multi-index $(x,\mu,a)$. 
We end up with the effective action
\begin{equation}
\mathcal{S}_{2}=\sum_{x}\left[\frac{2}{g}\left(1-\varphi^a_{x+\mu}\varphi^a_{x}-\frac{1}{2}A_{x\,,\mu}^{a\,2}\right)+\sum_y\psi^1_x\mathcal{K}_{x,y}\psi^2_y\mathcal{J}_{12}\right],
\label{final_eff_action}
\end{equation}
where
\begin{equation}
\mathcal{K}_{x,y}=N\,\delta_{xy}+\frac{2}{g}\left[\varphi^a_{x}\left(\varphi^a_{x+\mu}+\varphi^a_{x-\mu}\right)\delta_{xy}+(A_{x\,,\mu}^{a}+A_{x-\mu\,,\mu}^{a})\varphi^a_{x}\delta_{xy}-(\delta_{x-\mu,y}+\delta_{x+\mu,y})\right].
\end{equation}
Notice that $\mathcal{K}$ is symmetric under the exchange of $x$ and $y$.\\
We can finally integrate out the fermionic fields, which  leads to 
\begin{equation}
\mathcal{Z}=\int\prod_{x}dA_xd\varphi_xe^{-\mathcal{S}_{\text{bos}}}\;\det\mathcal{K}.
\label{final_path_int}
\end{equation}
 Since $\mathcal{K}$ is a real matrix, its determinant is real. However, we do not know a priori whether it is positive or not. It is then reasonable to expect the emergence of a sign problem in the simulations, an issue that will be analysed in the future. For the moment we have ignored the sign of the determinant, replacing $\det\mathcal{K}$ with its absolute value in \eqref{final_path_int}. We have then used the pseudofermion representation:
 \begin{equation}
 	|\det\mathcal{K}|=\sqrt{\det\mathcal{K}^2}\propto\int\,d\chi\exp\left(-\sum_{x,y}\chi_x^T(\mathcal{K}^2)^{-1}_{xy}\,\chi_y\right),
 \end{equation}
where the pseudofermion $\chi$ is real.

\noindent The final effective action that we have used for numerical simulations is then
\begin{equation}
\mathcal S_{\text{eff}}=\sum_x\,\frac{2}{g}\left(1-\varphi^a_{x+\mu}\varphi^a_{x}-\frac{1}{2}A_{x\,,\mu}^{a\,2}\right)+\sum_{x,y}\,\chi_x^T(\mathcal{K}^2)^{-1}_{xy}\,\chi_y.
\label{action_psuedo_f}
\end{equation}
\section{Simulation algorithm}
We have worked with a standard Hybrid Monte-Carlo~\cite{Duane:1987de}. We have chosen the Molecular Dynamics Hamiltonian 
\begin{equation}
\mathcal{H}=-\sum_{x}\left[\frac{1}{2}(\pi^a_x)^2+\frac{1}{2}(p^{a}_x)^2\right]+\mathcal{S}_{\text{eff}}(\varphi,A),
\end{equation}
where $\pi$ and $p$ are the conjugated momenta of $\varphi$ and $A$ respectively. The conjugated momentum $\pi_x$ is constrained to be orthogonal to $\varphi_x$, and this guarantees that $\varphi^2=1$ along the solutions of the equations of motion. Above, we omit the dependence on $\chi$ of $\mathcal{S}_{\text{eff}}$, since the pseudofermion is a spectator for the Molecular Dynamics. \\

\noindent The construction of symplectic integrators for this Hamiltonian is not entirely trivial due to the constraint on the bosonic field $\varphi^2=1$. A generalization of the leapfrong integrator reads
\begin{equation}
\begin{array}{l}
\pi^{a}_{1/2}=\pi^{a}_0-\frac{\tau}{2}(\mathcal{P}^{\varphi}_{0})^{ab}\frac{\partial \mathcal{S}_{\text{eff}}}{\partial \varphi^b}(\varphi_{0},A_0)\\
p^{a}_{1/2}=p^{a}_0-\frac{\tau}{2}\frac{\partial \mathcal{S}}{\partial A^a}(\varphi_{0},A_0)\\
\varphi^a_1=\cos(\tau|\pi_{1/2}|)\varphi^a_0+\sin(\tau|\pi_{1/2}|)\,\frac{\pi^{a}_{1/2}}{|\pi_{1/2}|}\\
A^a_1=A^a_0+\tau p^{a}_{1/2}\\
\pi^{a}_1=\cos(\tau|\pi_{1/2}|)\,\pi^{a}_{1/2}-\sin(\tau|\pi_{1/2}|)\,|\pi_{1/2}|\varphi^a_0-\frac{\tau}{2}\,(\mathcal{P}^{\varphi}_{1})^{ab}\,\frac{\partial \mathcal{S}_{\text{eff}}}{\partial \varphi^b}(\varphi_{1},A_1)\\
p^{a}_{1}=p^{a}_1-\frac{\tau}{2}\frac{\partial \mathcal{S}}{\partial A^a}(\varphi_{1},A_1).
\end{array}
\end{equation}
$\mathcal{P}^\varphi_x$ is the projector on the hyperplane perpendicular to $\varphi_x$
\begin{equation}
(\mathcal{P}^{\varphi}_x)^{ab}=\mathbbm{1}-\varphi^a_x\,\varphi^b_x,
\end{equation}
The momentum $p_x$ is generated from the Gaussian distribution $P(p)\propto e^{-p^{2}/2}$, while the momentum $\pi_x$ is constructed by generating an auxiliary momentum $\tilde\pi_x$ from the Gaussian distribution $P(\tilde\pi)\propto e^{-\tilde\pi^{,2}/2}$ and by setting $\pi_x=\mathcal{P}_x\tilde{\pi}_x$.

\section{Numerical explorations}
All simulations were run for $N=1$, $m=1$ and on a  $V=16\times 16$ lattice. The preliminary results presented here are obtained at three different values of the coupling $(g=0.1,g=1.0,g=10)$ and they all take into account the autocorrelations, computed with the particular version of the $\Gamma$-method described in~\cite{Wolff:2003sm}.

\begin{figure}[htp]
	\begin{subfigure}{0.5\textwidth}
		\includegraphics[width=\textwidth]{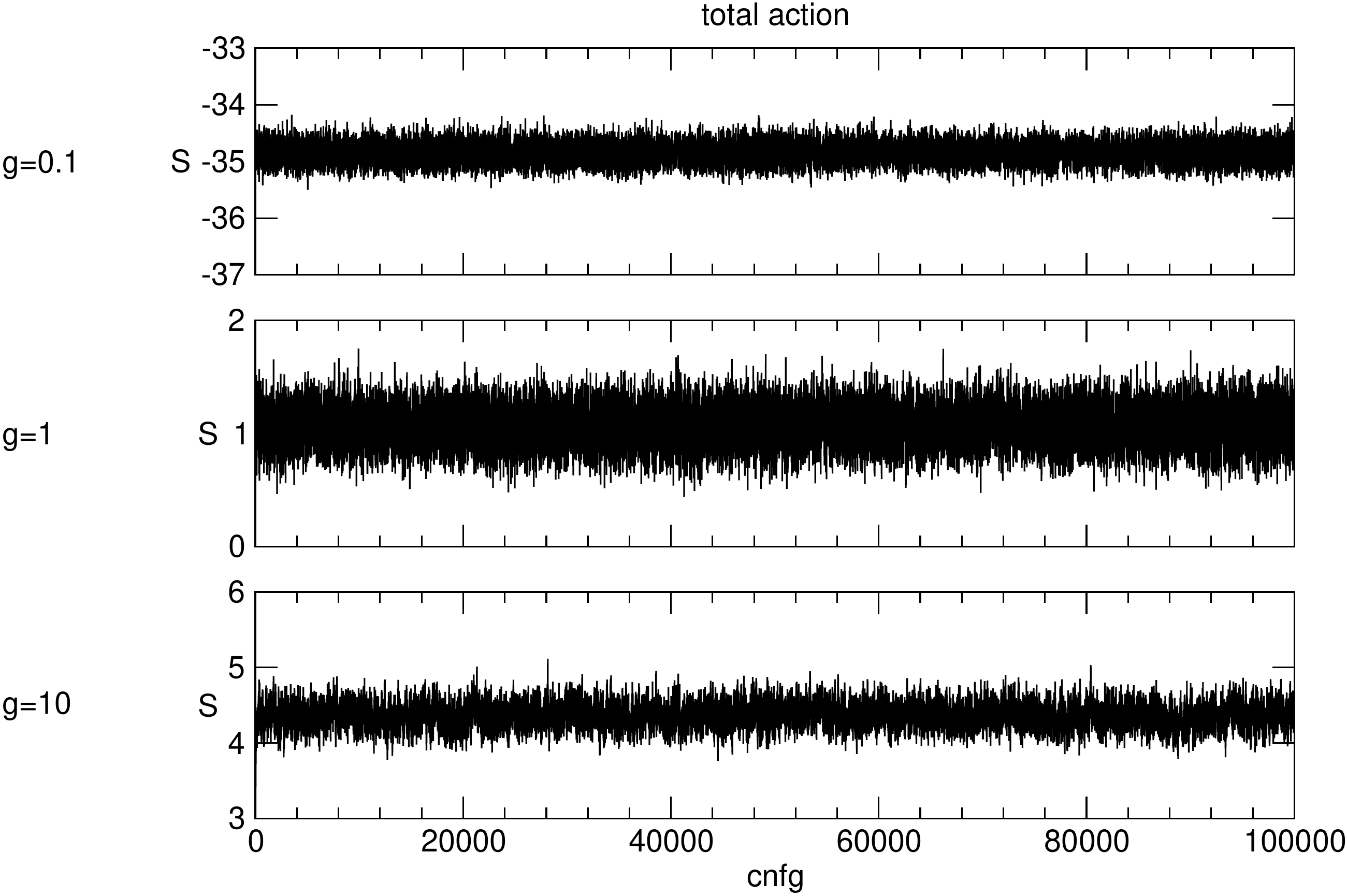}	
		\caption{Total action $\mathcal{S}_\text{eff}$ }
		\label{action}
	\end{subfigure}
	\begin{subfigure}{0.5\textwidth}
		\includegraphics[width=\textwidth]{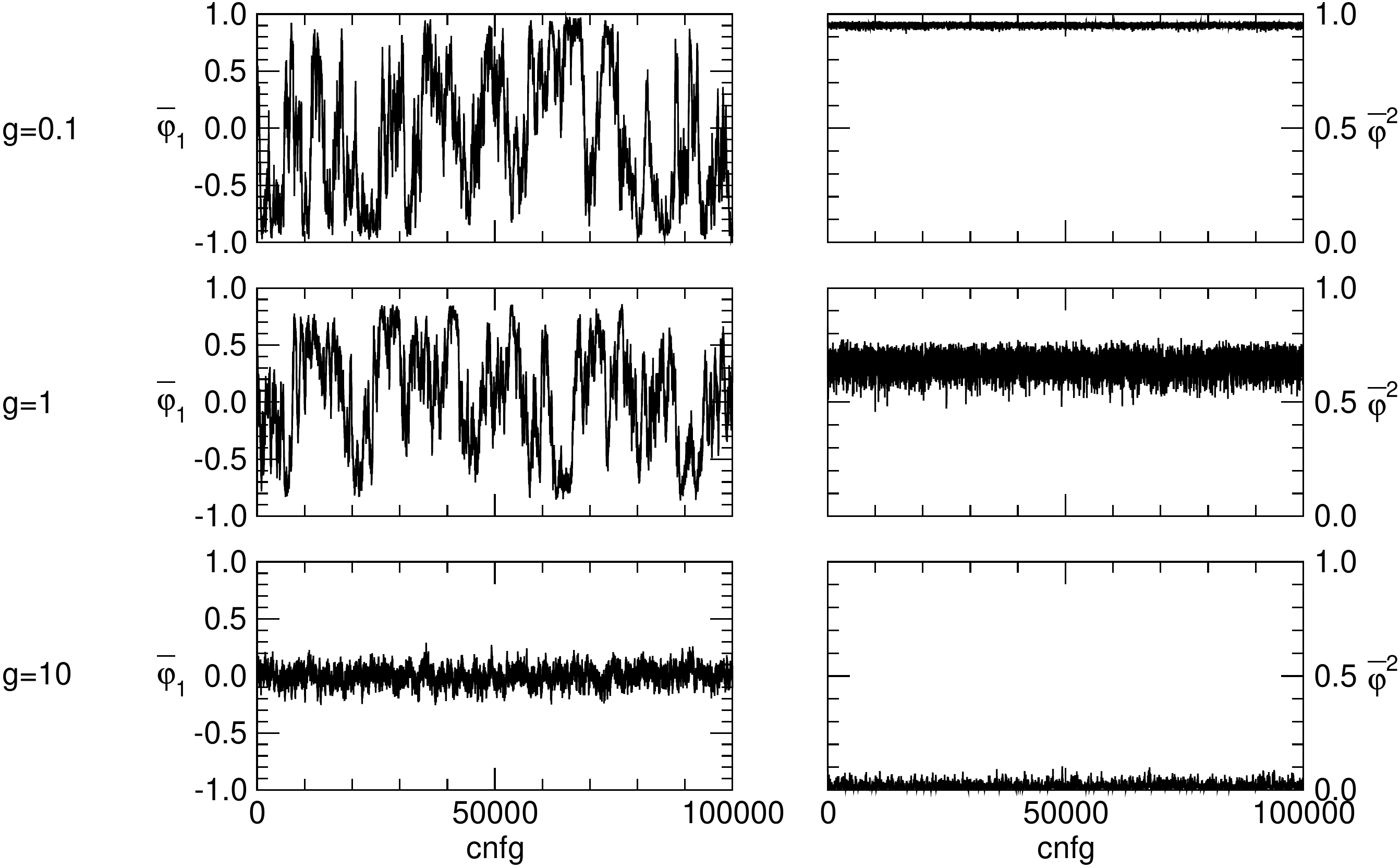}
		\caption{$\bar\varphi_1$ (left) and $\bar\varphi^2$ (right).}
		\label{mean_field}
	\end{subfigure}
\caption{History plots for the three different values of the coupling $g=0.1, g=1, g=10$.}
\label{hist_plot}
\end{figure}

\noindent Fig. \ref{action} represents the history plots of the total action \eqref{final_eff_action}. In Fig.~\ref{mean_field} we show the histories of two diagnostic observables $\bar{\varphi}_1=\frac{1}{V}\sum_{x}\varphi^1_x$ and $\bar{\varphi}^2=\sum_{a}(\frac{1}{V}\sum_{x}\varphi^a_{x})^2$.

\noindent Looking at the behavior of $\bar{\varphi}^2$ as a function of the three values of the coupling $g$, we see indication of a crossover between a symmetry-broken and an unbroken phase. 

\begin{figure}[htp]
	\centering
	\includegraphics[width=0.8\linewidth]{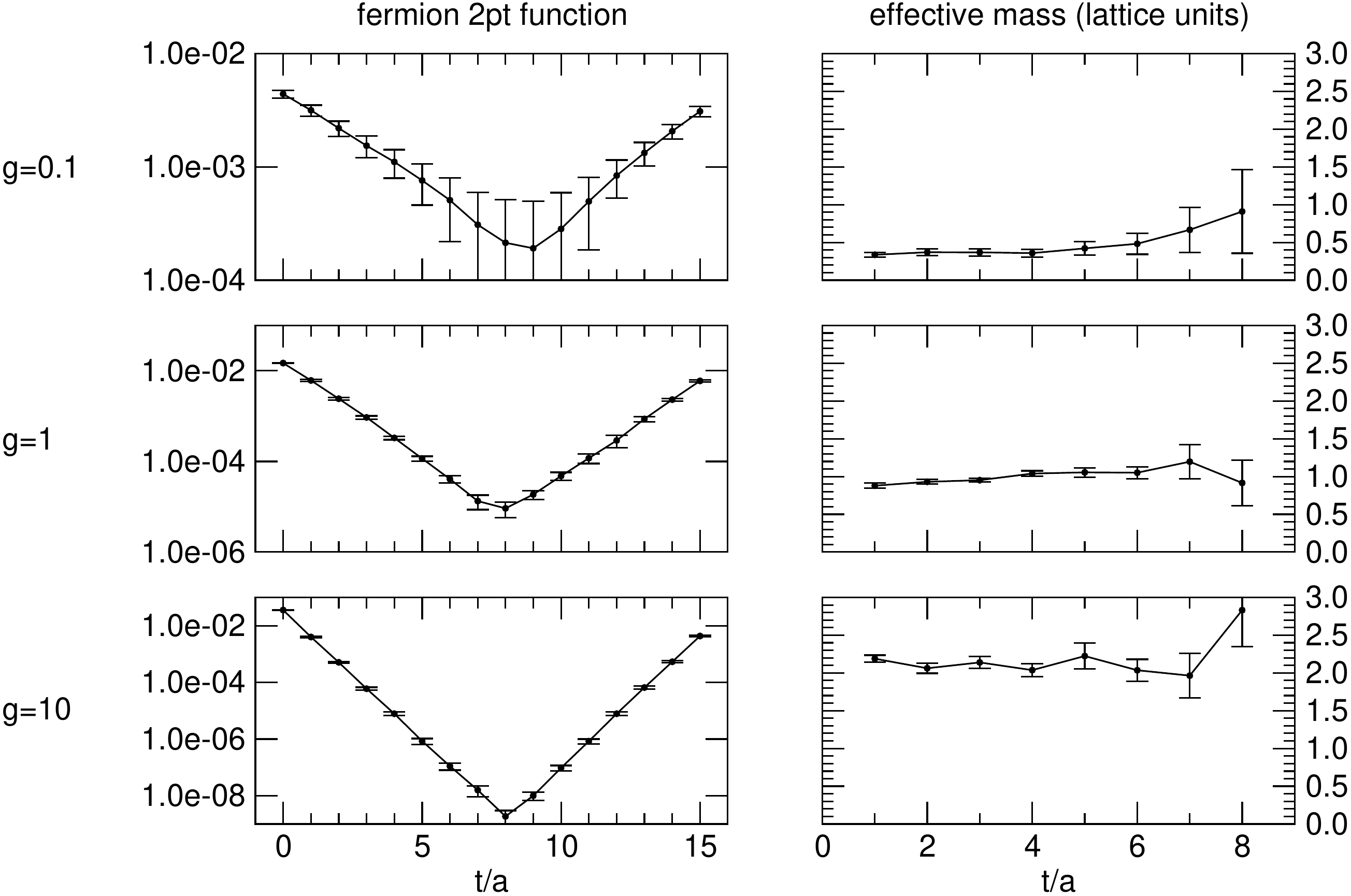}
	\caption{Fermion two-point function $C(t)$ (left) and the effective mass $M(t)$ (right) expressed in lattice units. Errors are only statistical.}
	\label{ffww}
\end{figure}

 \noindent Fig.~\ref{ffww} shows the fermionic two-point function $C(t)\equiv\sum_{x}\langle\psi^1(t,x)\psi^2(0)\rangle$ at the three values of $g$ and the effective mass $M(t)$, defined from the asymptotic form of the correlator
 \begin{equation}
 	C(t)\sim\, A\cosh\left[m\left(t-\frac{N_t}{2}\right)\right],
 \end{equation}
 by means of the equation
 \begin{equation}
 	\frac{C(t+1)}{C(t)}=\frac{\cosh\left[M(t)\left(t+1-\frac{N_t}{2}\right)\right]}{\cosh\left[M(t)\left(t-\frac{N_t}{2}\right)\right]}.
 \end{equation}
  We see that, as expected, the effective mass is large in the symmetric phase and small in the symmetry-broken phase. 

As mentioned above, these simulations ignore a potential sign fluctuation of the fermionic determinant. Its possible impact is not yet clear and will be object of future study.

\acknowledgments{The research of I.C. and T.M. is funded by the Deutsche Forschungsgemeinschaft (DFG, German Research Foundation) - Projektnummer 417533893/GRK2575 ”Rethinking Quantum Field Theory”. The research of V.F. is supported by the STFC grant ST/S005803/1 and ST/X000729/1, the European ITN grant No 813942 and from the Kolleg Mathematik Physik Berlin. The research of B.H. is supported by a UKRI Future Leaders Fellowship (grant number MR/T018909/1). }

\bibliographystyle{JHEP}
\bibliography{lattice22}
\end{document}